\documentclass[11pt]{article}

\usepackage[a4paper,margin=2.4cm]{geometry}
\usepackage{graphicx}
\usepackage{booktabs}
\usepackage{amsmath}
\usepackage{amssymb}
\usepackage{textcomp}
\usepackage{xcolor}
\usepackage{tikz}
\usetikzlibrary{positioning,arrows.meta}
\usepackage{hyperref}
\hypersetup{hidelinks}
\usepackage{natbib}
\usepackage{caption}
\usepackage{multirow}
\usepackage{authblk}
\usepackage{tabularx}
\usepackage{makecell}
\graphicspath{{figures/}}

\newcommand{\num}[1]{\textcolor{blue}{#1}}   

\title{An Ocean Model Ported by a Large Language Model: Experience and Lessons
from FESOM2 (Fortran to C to C++/Kokkos)}

\author[1]{N.~Koldunov}
\author[1]{S.~K.~Cheedela}
\author[1]{S.~Danilov}
\author[1]{D.~Sidorenko}
\author[1]{S.~Beyer}
\author[1,2]{T.~Jung}

\affil[1]{Alfred Wegener Institute, Helmholtz Centre for Polar and
          Marine Research, Bremerhaven, Germany}
\affil[2]{Department of Physics and Electrical Engineering, University of Bremen, Bremen, Germany}
\date{\today}

\begin{document}
\maketitle

\begin{abstract}

Large language models (LLMs) can translate and modify source code, and have been shown to do so for codes of different complexity. Whether they can port a complete, production geophysical model
to a different language without degrading its physics has not been established. We demonstrate that LLM-assisted code translation can preserve the physics of a complete production ocean model while moving it into a modern performance-portable form. We report our experience using an agentic LLM coding assistant, directed by domain experts, to port the FESOM2 unstructured mesh ocean--sea-ice model (about \num{74\,000} lines of core Fortran) first to C and then to C++/Kokkos for performance portability across CPUs and GPUs. We describe the practices that proved necessary, what worked and what did not, and the failure modes that we encountered. Three practices mattered most: translating in two stages that separate reproducing the numerics (Fortran to a clean C reference) from introducing parallelism (C to Kokkos); requiring a strictly literal translation in which the assistant was not permitted to ``improve'' the source; and validating each stage against an acceptance criterion suited to it. 

The C port reproduces the original Fortran at the level of long-term simulation statistics over five years. The Kokkos port is bit-for-bit identical to the C reference on CPU and statistically close on GPU over multi-year runs. On eddy-rich meshes up to \num{7.4}~million surface vertices a single A100 GPU node runs \num{1.6--3.7$\times$} faster than a CPU node, reaching the \num{1--2} simulated-years-per-day required for production integrations. The result is more than a single GPU port: by following a clear validation procedure, an LLM moved a full Fortran ocean model into another language and onto accelerators while preserving its physics in a matter of weeks.

\end{abstract}

\section{Introduction}
\label{sec:introduction}
Climate projections are moving toward kilometre-scale ocean resolution \citep{destine, moon2025earth, wedi2025, segura2025nextgems, John2025}, and that move requires addressing a lot of high-performance-computing problems. The established ocean general-circulation models are large Fortran codes, many of them developed over decades for distributed-memory CPU clusters. Fortran has served the field well, but it is increasingly less well supported by modern toolchains, libraries and hardware \citep{curcic2021toward}, and the gap is widest on the GPUs \citep{herten2023many}. Two related problems follow: keeping these models in a language and form that current and future systems support well, and running them on accelerators. Both have so far required years of specialist engineering, directive-based ports (OpenACC or OpenMP offload), domain-specific source-to-source tools, or complete rewrites in other languages. The scarce resource is therefore not only compute time, but the combination of human effort, domain knowledge, porting expertise, performance-tuning skill and validation of ports needed to make such translations scientifically trustworthy. This scarcity limits how quickly the community can move established models onto new hardware and build sustainable modeling infrastructure for the future.

Large language models (LLMs), and the agentic coding assistants built on them, can read and modify large codebases, and have demonstrably translated and modernized source code: across programming languages \citep{roziere2020transcoder}, in multi-step software-engineering tasks carried out under test feedback \citep{jimenez2024swebench}. More directly relevant to the present study, recent work has begun to examine LLM-assisted Fortran-to-C/C++ conversion, but so far at the scale of individual functions and small programs \citep{ranasinghe2025fortran,chen2024fortran2cpp}. 

The open question is therefore not whether an LLM can write plausible numerical code (it certainly can), but whether it can move a production geophysical model into a different programming language and framework, better supported by current and emerging hardware such as GPUs, without degrading the science.

In this work, the answer is affirmative. To our knowledge, this is the first demonstration that an LLM-assisted port can carry a full production ocean--sea-ice model to a GPU-capable implementation while retaining scientific fidelity and reaching production-relevant performance.

This paper reports how we achieved that result for FESOM2. Under the direction of domain experts, we used an agentic LLM coding assistant (Claude Code, running the Opus~4.7 model) to port the FESOM2 ocean--sea-ice model \citep{danilov2017fesom2}, an unstructured-mesh finite-volume model used in ocean and climate research, from its native Fortran to C and then to C++/Kokkos targeting both CPUs and GPUs. We present a documented account of what the task required and the working practices that proved necessary.

Rather than translating the Fortran directly to a parallel C++ target, we inserted an intermediate C stage. This was not just an implementation detail: by collapsing the highly configurable Fortran code onto the single configuration actually used in the run, the C stage resolved a major source of error before any GPU work began. In particular, it made explicit which compile-time options and runtime defaults were genuinely active — a question on which the assistant had frequently guessed incorrectly.

Each translation stage was validated against the acceptance criterion appropriate to it. The port of C is required to reproduce the original Fortran not bit-for-bit but at the level of long-term simulation statistics: over a five-year integration the global sea-surface temperature and salinity differ from the original by five-year mean root-mean-square differences of $0.006\,$\textdegree C and $0.002\,$ salinity units, with an interior temperature difference that is statistically indistinguishable from zero below 700\,m. We regard this as the appropriate bar for a change of language --- bit-for-bit agreement (within machine precision) across compilers and arithmetic orderings is neither attainable nor necessary --- whereas the move onto Kokkos is held to the stricter and achievable standard of bit-for-bit reproduction of the C reference on a deterministic (serial) back-end over a full simulated year. On the GPU, where reductions and scatters are reordered, the model is no longer bit-identical but remains statistically close over multi-year runs (annual sea-surface-temperature correlation $1.0$, bias $+10^{-4}\,$\textdegree C). On meshes with up to 7.4~million surface vertices --- eddy-rich configurations such as those used in nextGEMS \citep{segura2025nextgems} and DestinE \citep{destine,wedi2025} --- a single A100 GPU node runs $1.6$--$3.7\times$ faster than a CPU node, and the model reaches the 1--2 simulated-years-per-day band required for long production integrations. The same Kokkos code was also brought up on several other current accelerator systems --- JUPITER GH200, MN5 H100 and LUMI MI250X --- in each case requiring less than a day to compile and run the model, and we report first performance measurements on these platforms.

This paper makes three main contributions. First, it provides an account of porting a production ocean general-circulation model from Fortran to a modern, GPU-capable language using an LLM, together with an evaluation of its multi-year fidelity (Sections  ~\ref{sec:methods}, ~\ref{sec:fidelity}). Second, it identifies a set of failure modes specific to LLM-based translation of scientific code, paired in each case with the validation check that exposed the issue (Section ~\ref{sec:discussion} and Appendixes). Finally, it reports a direct performance comparison of the Fortran, C, Kokkos-CPU, and Kokkos-GPU implementations for four meshes spanning 0.13 to 7.4~million surface vertices on several GPU systems (Section \ref{sec:performance}). The appendices provide additional detail on the concrete porting steps and problems encountered in the C and Kokkos ports (\ref{sec:appendix-cport}, \ref{sec:appendix-kokkos}) They are intended in part as reusable context that can be supplied to an LLM when planning similar ports of other models.

\section{Related work}
\label{sec:relatedwork}

\paragraph{GPU and performance-portability ports of ocean and atmosphere models.}
Efforts to run ocean and atmosphere models on GPUs fall into three broad families. Directive-based ports annotate the existing Fortran with OpenACC or OpenMP-offload pragmas: examples include the GPU port of the ICON atmosphere \citep{giorgetta2022icongpu,lapillonne2026icongpu} and of Fortran ocean general-circulation models such as LICOM \citep{jiang2019licom}, while the PSyclone source-to-source approach has been applied to the NEMO ocean model \citep{porter2018nemolite2d}. These ports preserve the source language but tend to tie performance to a particular compiler, and several report mixed precision as a principal factor --- the operational ICON-on-GPU port attributes a large part of its speed-up to it \citep{lapillonne2026icongpu}. A related OpenACC effort in the framework of a natESM \citep{ehlert2025natesm} sprint ported FESOM2 tracer advection, sea-ice dynamics, and pressure- and density-dynamics kernels to GPUs and reported large kernel-level speed-ups, but the resulting implementation remained difficult to maintain and was not straightforward to bring up on other machines.

A second family rewrites the model in a C++ performance-portability framework: the E3SM atmosphere SCREAM/EAMxx \citep{caldwell2021scream,donahue2024eamxx} and the MPAS-Ocean \citep{ringler2013mpas} successor Omega \citep{petersen2026omega} are both built on Kokkos, obtaining single-source CPU/GPU portability at the cost of a full reimplementation. 

A third family writes the model outside the Fortran lineage entirely, as in Veros in Python/JAX \citep{hafner2018veros} and Oceananigans in
Julia \citep{ramadhan2020oceananigans}, inheriting GPU support from the host language. Our work is closest in spirit to the Kokkos rewrites, but differs in approach: rather than reimplementing the model from its equations, we port an existing, validated ocean model through a literal translation, and we automate that translation with an LLM.

\paragraph{Performance-portability layers.}
We use Kokkos \citep{trott2022kokkos,edwards2014kokkos}, which expresses parallel kernels and multidimensional data once and maps them to Serial, OpenMP, CUDA or HIP back-ends at build time. Alternatives include RAJA \citep{beckingsale2019raja} and SYCL \citep{khronos2020sycl}; the approach we describe is not specific to Kokkos and would apply to any of them.

\paragraph{LLMs and agentic tools for code translation.}
LLMs have been applied to programming-language translation since before the current generation of models \citep{roziere2020transcoder}, and agentic tools now perform multi-step software engineering under test feedback \citep{jimenez2024swebench}. The translation and modernization of legacy Fortran specifically has begun to attract attention, including LLM-assisted Fortran-to-C++ conversion \citep{ranasinghe2025fortran,chen2024fortran2cpp}. What remains largely absent from that literature is the translation of fidelity-critical numerical code, validated against the original at the scale of a full model, which is the niche of the present work.

\section{The FESOM2 model and the ported configuration}
\label{sec:model}
FESOM2 \citep{danilov2017fesom2} is a global ocean--sea-ice model that discretizes the ocean on an unstructured triangular surface mesh with a finite-volume scheme, using an arbitrary-Lagrangian--Eulerian vertical coordinate and a split of the fast barotropic and slow baroclinic dynamics. Scalar quantities (temperature, salinity, sea-surface height) live on mesh vertices while horizontal velocities live on the triangle elements; this vertices--elements staggering, together with the halo structure of its MPI parallelization, is a recurring source of confusion for LLM when porting. Such confusion is perhaps unsurprising: most ocean general-circulation models are formulated on structured or logically rectangular grids, so FESOM2's unstructured discretization, vertices--elements staggering and associated halo exchanges make it a less conventional target for LLM-assisted translation. We refer the reader to \citet{danilov2017fesom2,scholz2019fesom2} for the model formulation and concentrate here on what was ported.

The ported configuration use the subset of FESOM2 options, that is comparable to the standard ocean--ice setups used by many ocean models in OMIP2 \citep{tsujino2020evaluation} and in CMIP6 \citep{eyring2016overview}. It comprises the linear-free-surface treatment of sea surface height, the full nonlinear equation of state, the hydrostatic pressure-gradient force, second-order Adams--Bashforth time stepping, flux-corrected tracer advection (with high-order horizontal and vertical schemes), biharmonic momentum viscosity, both the simple Pacanowski--Philander (PP) and the K-profile vertical-mixing (KPP) schemes (the latter the default), the Gent--McWilliams/Redi eddy parameterization, a conjugate-gradient sea-surface-height solver, elastic--viscous--plastic sea-ice dynamics with thermodynamics, and atmospheric forcing through bulk formulae with sea-surface-salinity restoring and runoff.

This scoping matters for the code-size figures. The core ocean--ice Fortran actually exercised by the configuration is 109 source files totaling about 74{,}000 lines. The full FESOM2 distribution, which adds the biogeochemistry and the external mixing library, sea-ice thermodynamics, and libraries for coupling with atmosphere, as well as multiple I/O options, is 211 files and about 269{,}000 lines, none of which the configuration used in this work requires. The ports that reproduce this core are about 20{,}000 lines of C and about 31{,}000 lines of C++/Kokkos. The four meshes used in the performance evaluation, ranging from 0.13 (low resolution, 1-degree equivalent) to 7.4 million surface vertices (eddy-resolving), are listed in Table~\ref{tab:meshes}.

\begin{table}[t]\centering
\caption{Meshes used in the performance evaluation.}
\label{tab:meshes}

\begin{tabularx}{\textwidth}{lrrrrrX}
\hline
Mesh &
\makecell{surface\\vertices (M)} &
levels &
\makecell{$\Delta t_\mathrm{meas}$\\(s)} &
\makecell{$\Delta t_\mathrm{prod}$\\(s)} &
\makecell{Resolution\\(km)} &
Mesh class \\
\hline
CORE2 & 0.13 & 48 & 1800 & 1800 & 100--25 & 1\textdegree  \ equivalent \\
fArc  & 0.64 & 48 & 900  & 900  & 100--4.5 & regional, Arctic-refined \\
DARS  & 3.2  & 57 & 180  & 240  & 25--4 & eddy-resolving in energetic regions \\
NG5   & 7.4  & 70 & 180  & 240  & 14--5 & eddy-resolving over most of the globe \\
\hline
\end{tabularx}

\end{table}

\section{The porting process and its validation}
\label{sec:methods}

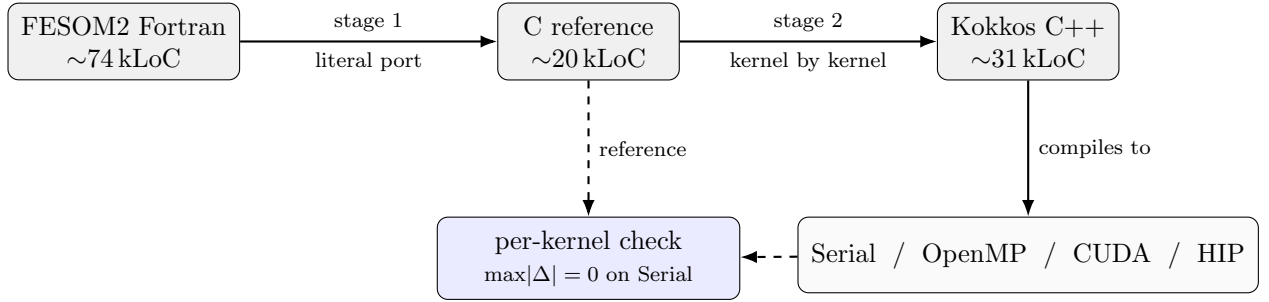
\begin{figure}[t]\centering
\begin{tikzpicture}[font=\small,
  box/.style={draw, rounded corners, align=center, inner sep=5pt, minimum height=10mm, minimum width=24mm},
  src/.style={box, fill=gray!12},
  be/.style={box, fill=gray!4, minimum width=56mm},
  gate/.style={box, fill=blue!8, minimum width=40mm},
  arr/.style={-{Latex[length=2.2mm]}, thick},
  lab/.style={font=\scriptsize, align=center}]
  \node[src] (f) {FESOM2 Fortran\\$\sim$74\,kLoC};
  \node[src, right=34mm of f] (c) {C reference\\$\sim$20\,kLoC};
  \node[src, right=34mm of c] (k) {Kokkos C++\\$\sim$31\,kLoC};
  \draw[arr] (f) -- node[lab, above]{stage 1} node[lab, below]{literal port} (c);
  \draw[arr] (c) -- node[lab, above]{stage 2} node[lab, below]{kernel by kernel} (k);
  \node[be, below=18mm of k] (back) {Serial \;/\; OpenMP \;/\; CUDA \;/\; HIP};
  \node[gate, below=18mm of c] (g) {per-kernel check\\{\scriptsize max$|\Delta|=0$ on Serial}};
  \draw[arr] (k) -- node[lab, right]{compiles to} (back);
  \draw[arr, dashed] (c) -- node[lab, right]{reference} (g);
  \draw[arr, dashed] (back) -- (g);
\end{tikzpicture}
\caption{The two-stage porting workflow. Fortran is translated to a clean C reference (stage~1) and
then wrapped in C++/Kokkos (stage~2); the C reference survives as a bit-identical Serial build
against which every Kokkos kernel is checked.}
\label{fig:method}
\end{figure}

\begin{figure}[t]\centering
\begin{tikzpicture}[font=\small,
  rung/.style={draw, rounded corners, text width=7cm, align=left, inner sep=5pt, minimum height=7mm},
  arr/.style={-{Latex[length=2mm]}, thick}]
  \node[rung, fill=green!12] (r1) {\textbf{R1} per-kernel (Serial/OpenMP): max$|\Delta|=0$};
  \node[rung, fill=green!12, above=2.5mm of r1] (r2) {\textbf{R2} whole model (Serial): byte-identical, 1 model year};
  \node[rung, fill=green!8, above=2.5mm of r2] (r3) {\textbf{R3} threaded (OpenMP): $=$Serial (maps); $\lesssim10^{-12}$ (scatters)};
  \node[rung, fill=orange!18, above=2.5mm of r3] (r4) {\textbf{R4} GPU (CUDA): statistically close, 20-step active-ice gate};
  \node[rung, fill=orange!18, above=2.5mm of r4] (r5) {\textbf{R5} multi-year parity: corr$\approx$1, drift$\approx$0};
  \draw[arr] (r1)--(r2); \draw[arr] (r2)--(r3); \draw[arr] (r3)--(r4); \draw[arr] (r4)--(r5);
\end{tikzpicture}
\caption{The validation ladder; green tiers are bit-identical, orange tiers statistically close. A
regression at any tier is caught before it can reach the science.}
\label{fig:ladder}
\end{figure}
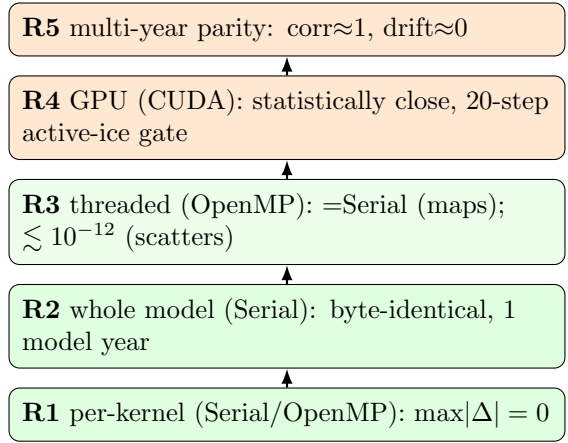

This section describes how we performed the port and how we checked it at each step. Three working practices organize what follows. These practices emerged during the work and were sharpened by an earlier failed attempt. First, we translated the model in two stages, separating the numerics from the parallelism. We required a literal, line-by-line translation that gave the assistant no room to ``improve'' the code, so that any difference from the reference was, by construction, a bug. And we checked each stage against a reference at the strictest level it could meet. Figure~\ref{fig:method} sketches the workflow and Figure~\ref{fig:ladder} the tiers of checking, which we refer to below as the validation ladder. Two appendices collect further details specific to the FESOM2 port (Appendices~\ref{sec:appendix-cport} and~\ref{sec:appendix-kokkos}); we include them less as instructions for human readers than as context that could be given directly to a language model in a future port.

\subsection{Two stages: Fortran $\rightarrow$ C $\rightarrow$ Kokkos}
Porting Fortran directly to a GPU programming model mixes two independent sources of error: mistranslating the numerics, and mis-parallelizing them. We separated them. The first stage produces a clean, single-threaded C reference, free of Fortran's module-global state (every \texttt{USE}-global becomes an explicit field of a passed struct), which exposes the data flow and is straightforward to reason about. The second stage wraps that C in C++/Kokkos for performance portability. What makes the second stage safe is that the C$\rightarrow$C++ transliteration changes no arithmetic: compiled at matched flags, the C++ build is bit-for-bit identical to the C build. The C code therefore survives, untouched, inside the Kokkos tree as a single-threaded reference (Section ~\ref{sec:ladder}), and kernels can be moved onto the device one at a time while the model stays runnable and verifiable at every commit.

We chose Kokkos~\citep{trott2022kokkos,edwards2014kokkos} as the portability layer because it is single-source (one C++ code compiles to Serial, OpenMP, CUDA or HIP back-ends by build configuration) and vendor-neutral: development used NVIDIA A100 GPUs, but we were able to run the same source on other machines, including ones with AMD GPUs without a rewrite (Section ~ \ref{sec:performance}). A shell script
enforces that no raw CUDA enters the source.

A further practical reason for choosing C as the intermediate language is that it moves the assistant into a higher-resource programming-language regime. Code LLMs are known to perform best on languages that are well represented in their training data and to degrade on low-resource or niche programming languages (Fortran in this case). For our purposes, C is therefore not only a convenient systems language but also a stabilizing intermediate representation: it gives different LLMs a better-supported source than Fortran, and it makes subsequent C→C++/Kokkos transformation closer to the language families on which coding models are the strongest.

A less obvious benefit of the intermediate C step is that it forces the model to be pinned down. FESOM2, like any mature ocean model, is a large configurable superset: a given run is selected from many compile-time switches and namelist options on top of the defaults written into the code, and only a fraction of the source is live for any one experiment. A porter --- human or LLM --- must therefore keep deciding which option is active and which value actually takes effect, and the LLM assistant is especially prone to following a plausible but inactive branch or adopting a module's hardcoded default. The C step removes that ambiguity: it commits to the single configuration under study.

Finally, the two stages call for different acceptance criteria, a point we return to in Section ~\ref{sec:fidelity}. The Fortran-to-C stage is accepted by statistical agreement over multi-year runs, not by bit-for-bit equality.

Bit-for-bit equality is reserved for the C-to-Kokkos stage, where the C reference and the Kokkos Serial back-end share a compiler, so that exact equality is both achievable and a precise detector of parallelization errors.

\subsection{Literal, line-by-line translation}
The single most important instruction to the assistant was to port the Fortran line by line and never to simplify, approximate or ``improve'' it. The only changes allowed are mechanical: 1-based to 0-based indexing; Fortran's column-major to C's row-major storage (through fixed index macros); and \texttt{USE}-globals to struct passing. Nothing else.

This rule is not about style; it is what makes the port debuggable. Because the Fortran is the verified specification and is stable at the production time step, any divergence of the port from the Fortran's behavior is a port bug by definition. That single rule narrows the debugging search: there is never a question of whether the physics is ``supposed to'' behave differently. The first
author had attempted this port before, and that attempt failed precisely because small ``improvements'' accumulated --- linearizing the thermal and haline expansion coefficients (which broke the surface boundary-layer depth), replacing a KPP lookup table with a closed-form approximation (wrong magnitudes), or dividing by a volume ratio instead of multiplying by an area ratio (which left the mesh geometry visible in the tracer field) --- each of which produced plausible output and a long debugging effort. The literal-translation rule existed to make that kind of shortcut impossible.
An alternative would be to verify translated functions through unit testing. However, many scientific codes rely heavily on non-local state, implicit dependencies, and side effects, making individual procedures difficult to test in isolation. As a result, establishing functional equivalence at the procedure level is often impractical.

We added one further rule, for constants: any statement about what the Fortran does with a constant must quote the exact \texttt{file:line} and the literal value read there, never one recalled, rounded, or copied from a comment. We adopted it after a constant was reported from a comment that confidently stated the wrong value: the value was trusted without being checked against the code, and a numerical instability stayed hidden for several working sessions until the rule forced LLM to read the actual line.

\subsection{Claude Code Workflow}
The work was carried out with Claude Code, an agentic command-line coding assistant, driving the Claude Opus~4.7 model at \texttt{max} reasoning effort and running in auto-accept mode, on the DKRZ Levante system. A port of this size cannot be completed in one sitting; ours ran across many working sessions, and the main practical difficulty was to carry the project's state from one session to the next. Claude Code does part of this on its own: it reads a project file (\texttt{CLAUDE.md}) at the start of a session and keeps a small per-project memory that it recalls later. For a port of this length, however, that built-in memory was not enough on its own --- partly because there is simply too much to carry, and partly because a session whose context grows very large becomes slower and less reliable, so it is better to keep each session short and narrowly scoped than to let one run accumulate the whole history.

We therefore relied on a few file-based practices, built on the open ``cc-thingz'' skills for Claude Code (\url{https://github.com/umputun/cc-thingz}). Before each major step we used the \emph{brainstorming} skill to fix the direction and the focus of the coming phase, and then the \emph{planning} skill, which writes the plan to a file on disk. This plan file captures everything a later session needs to rebuild context and it tracks progress through the phase (each completed milestone git-tagged), so that the individual subtasks can each be done in a separate short, low-context session that starts from the plan rather than from the full conversation. Two further running files supported it: a lessons-learned file, to which each session added the mistakes and their fixes so they were not repeated, and a handoff file written at the end of every session recording the current state, the canonical reference run, and the next task. Because all of these are ordinary files kept with the code, a fresh session could reconstruct what it needed from a few reads and begin work at once.

The division of labour was stable throughout. The human acted as ocean-model domain expert and director --- setting strategy, reviewing plans and diffs, authorizing the few deliberate departures from a literal port (the I/O subsystem was redesigned rather than transliterated), and catching the subtle physics and validation errors --- while the assistant did the reading, translation, harness construction, debugging, profiling and documentation. Table~\ref{tab:effort} summarizes the scale and effort.

\begin{table}[t]\centering
\caption{Code size and porting effort. The ``Fortran (core)'' column is the top-level
\texttt{src/*.F90} actually exercised by the ported configuration; the full FESOM2 distribution (adding REcoM biogeochemistry, CVMix, Icepack and IFS coupling, none of which the configuration requires) is 211 files and 268{,}910 lines.}
\label{tab:effort}
\begin{tabular}{lrrr}
\toprule
                 & Fortran (core) & C port            & Kokkos port \\
\midrule
language         & Fortran        & C                 & C++/Kokkos \\
files            & 109            & 73                & 78 \\
lines            & 73{,}991       & 20{,}010          & 30{,}663 \\
days with commits & ---            & 10                & 7 \\
\bottomrule
\end{tabular}

\end{table}

\subsection{The Kokkos validation ladder}
  \label{sec:ladder}
This ladder validates the second stage of the port: the Kokkos rewrite is checked against the C reference from the first stage, tier by tier, from a single kernel up to a multi-year run. Keeping the C reference inside the Kokkos tree, built with the same compiler, is what lets the acceptance be exact on the CPU back-ends --- a bit-for-bit match that the Fortran-to-C stage, crossing both languages and compilers, could never provide. The C reference's own agreement with the original Fortran is the separate, statistical question of Section ~\ref{sec:fidelity}. The tools that locate a failure, as opposed to the gates here that detect one, were built during the C port and reused for Kokkos (Section ~\ref{sec:tools}).

Its lowest tier checks one Kokkos kernel at a time. When a kernel is rewritten in Kokkos, the original C version of that same kernel is left in the source as a reference, which we call its ``twin''. An optional check, switched on with an environment variable (\texttt{FESOM\_KK\_VERIFY}), then has the running model do the following each time the kernel is called: save the new Kokkos output, run the C twin on the same live model state, record the largest absolute difference between the two, and put the Kokkos output back so the run proceeds unchanged. Because it restores its own result the check disturbs nothing and can be left on for an entire production run, and on the Serial Kokkos back-end the two must agree exactly --- the acceptance threshold is a maximum difference of zero. It tests more than the loop: rewriting a kernel also means copying its numerical constants into the new code and because the twin keeps its own copy, a single mistyped coefficient makes the two disagree and the check fails.

The higher tiers widen the scope. On Serial back-end, the whole model must stay byte-identical to the C reference over a full simulated year of the realistic configuration (the CORE2 mesh, 17{,}280 time steps at a fixed rank count, every field in all thirteen monthly snapshots). The OpenMP back-end is held to the same exact standard for kernels that only map or gather, and to a small floor ($\lesssim 10^{-12}$ per step) for the few that sum across threads, where the order of floating-point addition changes. The GPU back-end has a mandatory short gate before each commit: a twenty-step run of the realistic configuration with active sea ice, compared field by field against the Serial reference with per-field ceilings; a real bug saturates this metric about a thousand times above the floor, so the gate cleanly separates expected last-bit GPU divergence from an error. The top tier is multi-year agreement against both the Fortran and the C references, reported in Section ~\ref{sec:fidelity}.

Bit-identity holds only at a fixed MPI rank count, because both the halo-exchange order and the order of the conjugate-gradient solver's global sums depend on how the mesh is divided.

\subsection{Differential-testing and debugging tools}
\label{sec:tools}
The ladder tells us that a tier has \emph{failed}, but we also need to understand \emph{where}. Each of the following debugging tools compares the port against the reference, at a finer grain than the tier above it. We built them by instrumenting both sides of the comparison, the original Fortran included. They are small modules: some dump state, some check ranges, some switch a subsystem off. Each sits behind an
environment variable and is compiled out when that variable is unset, so with the instrumentation off the reference is byte-for-byte unchanged. We built all of them for the C port and reused them, unchanged, for Kokkos. 

\paragraph{Per-substep reference dump.} The ocean time step runs as about seventeen substeps. After each one, a dump module records a few state variables at a handful of probe vertices after each substep, written as raw little-endian binary (each record tagged with step, substep, vertice and field name) in a layout the C port reads back exactly. A short script then reports the first substep where the port and the reference disagree. On a chaotic system, ``the run diverges by step two'' says nothing about the cause, this names the guilty substep instead. The sea-ice, vertical-mixing and start-up routines iterate internally, so each one got its own dump points.

\paragraph{Identical-input operator diff.} A whole-run difference cannot tell a wrong kernel from a correct kernel that was fed a slightly different input, because both diverge. This tool removes the input difference. It feeds a prescribed input field, the same on both sides by construction, runs each kernel once, and compares the outputs point by point. A replay variant handles kernels that sit just after a step that magnifies tiny input differences. It overwrites the port's inputs with the reference's dumped inputs immediately before the routine, leaving only arithmetic. 

\paragraph{Always-on sanity probes.} For every field with a known physical range, a one-line check writes to standard error whenever a value leaves that range.

\paragraph{Stale-halo probe.} The most common distributed-memory fault is a stale halo: a field whose write loop did not cover its halo points. The probe finds it directly. Copy the field, exchange the copy, and compare it with the original over the halo range. Any point that changed was stale.

\paragraph{Subsystem disable-switches.} Each subsystem has an environment switch that turns it into a no-op. We can freeze the tracers, zero the wind or the heat flux, skip tracer advection or diffusion, disable the sea-ice dynamics or thermodynamics, or turn off the eddy parameterization. The switches compose, so three or four short runs bracket a multi-rank divergence to a single subsystem before any code is read. They also serve as bit-identity off-switches at a phase boundary. With the new physics switched off, the model must reproduce the previous commit byte for byte. That confirms the new code is inert when it should be, and it catches gating mistakes.

\section{Fidelity of the ports}
  \label{sec:fidelity}
Fidelity follows the two stages of the port, and each stage is held to the acceptance bar that fits it. The first stage, Fortran to C, cannot be bit-for-bit: a change of language, compiler and arithmetic ordering makes byte-level equality neither attainable nor meaningful, so the C port is judged by whether it reproduces the original model's simulated ocean over multi-year runs. The second stage, C to Kokkos, is held to the stricter and achievable standard of byte-level identity against the C reference on a deterministic back-end. Because that identity holds, the multi-year agreement established for the C port carries over to the Kokkos CPU build unchanged, and the only thing the GPU introduces that must be measured is the small divergence from reordered floating-point arithmetic. We take the two stages in that order.

\subsection{Stage 1: the C port reproduces the Fortran over multi-year runs}
\begin{figure*}[t]\centering
  \includegraphics[width=\linewidth]{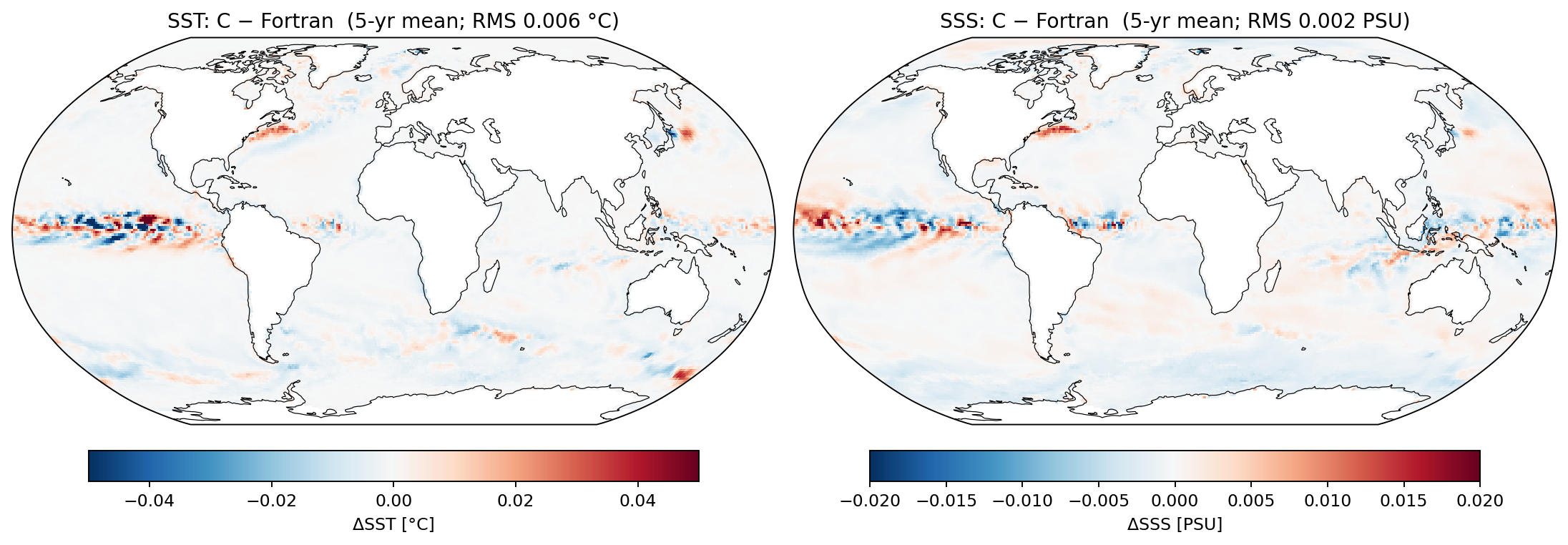}
  \caption{Five-year-mean (1958--1962) difference, C port minus the Fortran reference, in sea-surface temperature (left) and salinity (right). The differences have an area-weighted root-mean-square of $0.006\,$\textdegree C (SST) and $0.002\,$PSU (SSS). Both runs use identical physics, forcing and time step, so the difference is the accumulated Fortran$\rightarrow$C language-port divergence over five model years.}
  \label{fig:climate}
  \end{figure*}

\begin{figure*}[t]\centering
  \includegraphics[width=0.95\linewidth]{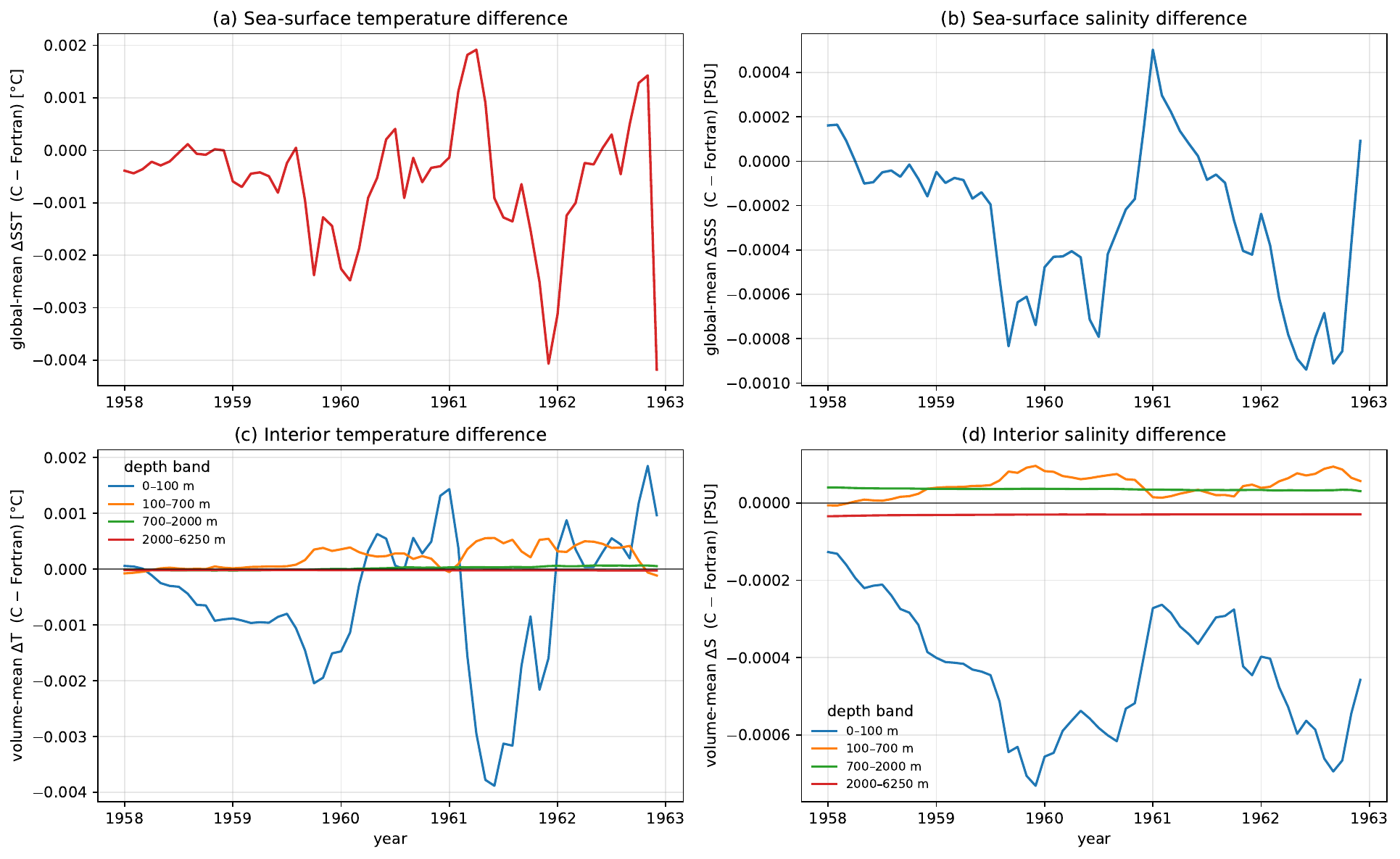}
  \caption{Five-year (1958--1962) difference, C port minus the Fortran reference. (a)~global-mean sea-surface temperature and (b)~salinity, monthly; (c)~volume-weighted interior temperature difference and (d)~salinity difference, each in four depth bands. The surface differences stay within a few thousandths of a degree and of a salinity unit and do not grow, and both interior differences are statistically zero below $700\,$m, confirming no deep-ocean drift between the runs.}
  \label{fig:drift}
  \end{figure*}

The C port is required to reproduce the original Fortran not bit-for-bit but at the level of long-term simulation statistics. We established this against the original Fortran model running the same configuration (CORE2 mesh), with the same forcing from 1958 onward at the production time step. Over five years the C port tracks the matched Fortran run with five-year-mean sea-surface temperature and salinity root-mean-square differences of $0.006\,$\textdegree C and $0.002\,$PSU (Fig.~\ref{fig:climate}), the residual structure confined to the western boundary currents, the Antarctic Circumpolar Current and Equatorial area of strongest variability rather than any basin-scale bias. Both codes share the same physical surface-temperature drift of about $-0.13\,$\textdegree C over the period, and their differential drift stays below $0.001\,$\textdegree C; the volume-weighted interior temperature and salinity differences are statistically zero below 700\,m (Fig.~\ref{fig:drift}), so the deep ocean shows no drift between the two runs. The Sea-ice area and volume in both hemispheres also track the Fortran run (Table~\ref{tab:fidelity}). The biggest difference is the sea-ice velocity, that correlates with the Fortran reference at only $\sim$$0.92$.

  \subsection{Stage 2: the Kokkos port reproduces the C reference}

  \paragraph{Bit-identity on CPU.} On a deterministic back-end the Kokkos port reproduces its C reference exactly. Every kernel ported to the device path passed its per-kernel gate (Section ~\ref{sec:ladder}) at a maximum absolute difference of zero against the C twin on the Serial back-end. At the whole-model level, each Kokkos back-end reproduces the C reference byte-for-byte over a complete simulated year of the realistic configuration (CORE2 mesh, 17{,}280 steps at 256 ranks, every field in all thirteen monthly snapshots), and the OpenMP back-end is byte-identical to Serial for all map and gather kernels. This is identity against the C port, not against the Fortran; but because the C port already matches the Fortran over multi-year runs (Stage 1), the Kokkos CPU build inherits that multi-year fidelity unchanged.

\paragraph{The GPU stays statistically close.} On the GPU the model is no longer bit-identical, but the residual difference is small and has two harmless causes, both expected. The first is rounding: the GPU performs some arithmetic --- a combined multiply-and-add, and the built-in exponential and logarithm used in the equation of state --- in a different order from the CPU, so the two results differ only in the last bit. The second is summation order: where the model adds many numbers in parallel, the additions happen in a different order than on a single core, and a floating-point sum depends slightly on that order. Together they set a steady floor of about $10^{-3}$ per step for temperature and $10^{-4}$ for velocity, and we record this fixed pattern so that a real regression stands out against it.

Table~\ref{tab:fidelity} reports the one-year comparison of the CUDA port against both references. Against the C port --- which, being bit-identical to Serial Kokkos, isolates this pure GPU rounding and summation drift --- the sea-surface temperature correlation is $1.0$ with a bias of $+10^{-4}\,$\textdegree C, salinity correlation $0.99996$, and sea-ice concentration and thickness correlations $0.99997$ and $0.99998$. The GPU drift is thus about three orders of
magnitude below the model's own C-versus-Fortran uncertainty from Stage~1; composing the two stages, the GPU reproduces the Fortran's multi-year statistics to within that same agreement.

  \begin{table}[t]\centering
  \caption{One-year fidelity of the Kokkos-CUDA port against the Fortran reference and against the C port (the latter isolating pure GPU reduction/scatter drift, since the C port is bit-identical to Serial Kokkos). Correlation and area-weighted bias; bias units are \textdegree C (SST), PSU (SSS), m (SSH), fraction (a\_ice), m (m\_ice), m\,s$^{-1}$ (uice).}
  \label{tab:fidelity}
\begin{tabular}{lcccc}
\toprule
       & \multicolumn{2}{c}{CUDA vs Fortran} & \multicolumn{2}{c}{CUDA vs C port} \\
 field & corr & bias & corr & bias \\
\midrule
SST    & 1.0 & $+3.1\times10^{-5}$ & 1.0 & $+1.0\times10^{-4}$ \\
SSS    & 0.99996 & $-5.3\times10^{-4}$ & 0.99996 & $-1.8\times10^{-4}$ \\
SSH    & 1.0 & $+2.0\times10^{-5}$ & 1.0 & $-1.4\times10^{-5}$ \\
a\_ice & 0.99997 & $+1.6\times10^{-4}$ & 0.99997 & $+1.6\times10^{-4}$ \\
m\_ice & 0.99998 & $-1.5\times10^{-4}$ & 0.99998 & $-1.5\times10^{-4}$ \\
uice   & 0.919   & $-1.0\times10^{-3}$ & 0.99978 & $-7.5\times10^{-5}$ \\
\bottomrule
\end{tabular}

  \end{table}

\section{Performance, scaling and portability}
  \label{sec:performance}
We did not set out to produce the fastest possible C version of FESOM2, nor the most highly tuned Kokkos port; either would call for considerable further engineering beyond the scope of this work. The question we ask is twofold. First, whether the model the LLM produced is performance-competitive from the outset --- fast enough, on both CPU and GPU, to be useful for real simulation rather than merely correct. Second, whether it is genuinely portable: whether the single Kokkos source runs unchanged across machines, and across GPU vendors, rather than only on the hardware it was written on. The numbers below are meant as evidence of a competitive and portable starting point, not of an optimization ceiling: where the port already keeps pace with, or exceeds, the original Fortran, a deliberate tuning effort would have room to improve on it further.

\begin{figure*}[t]\centering
\includegraphics[width=0.9\linewidth]{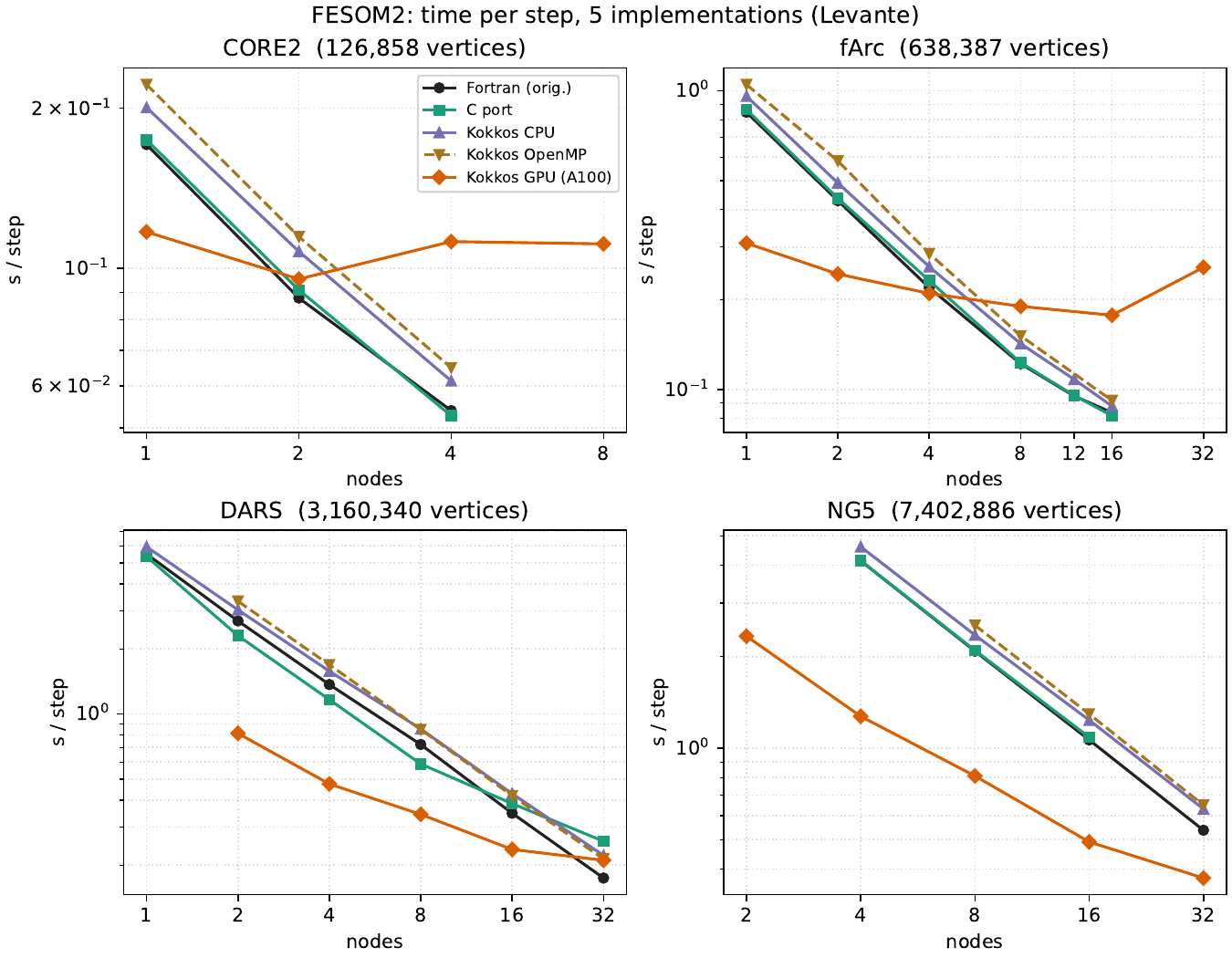}
\caption{Time per step for the five implementations across the four meshes (log--log.}
\label{fig:compare}
\end{figure*}

\begin{figure}[t]\centering
\includegraphics[width=\linewidth]{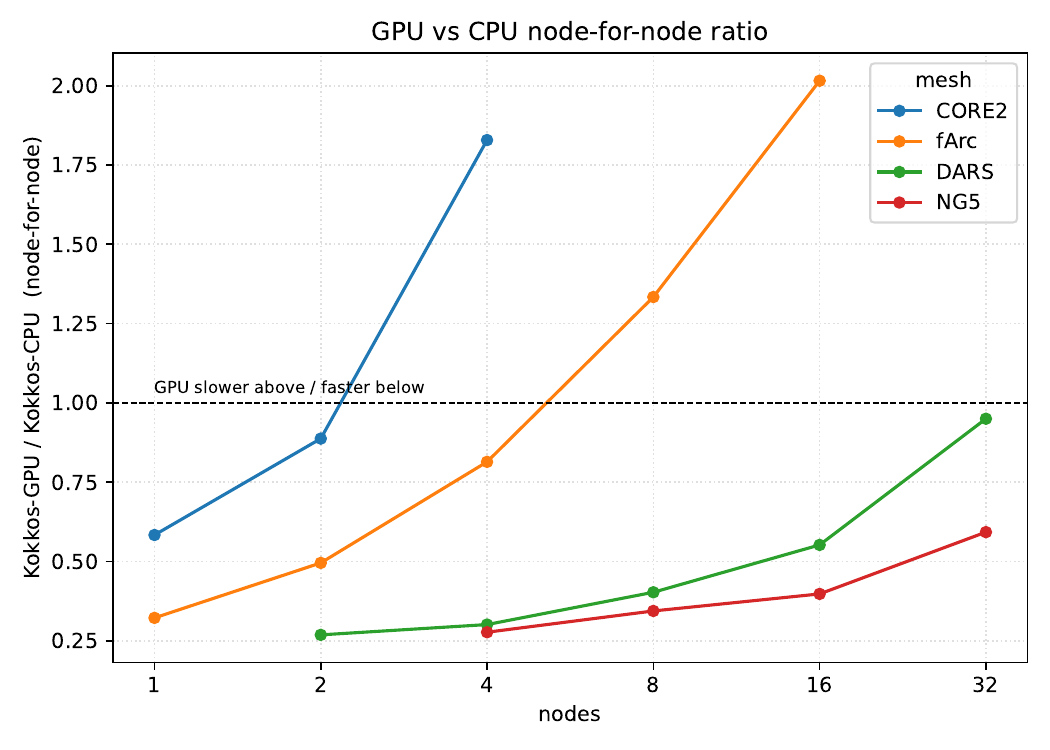}
\caption{Node-for-node Kokkos-GPU / Kokkos-CPU step-time ratio (below one means the GPU is faster). The GPU advantage is largest at low node count and erodes as subdomains thin.}
\label{fig:ratio}
\end{figure}

The same Kokkos source compiles and runs on NVIDIA GPUs through the CUDA back-end and on AMD GPUs through the HIP back-end, as well as on CPUs through the Serial and OpenMP back-ends, with no change to the physics code. Most of the measurements in this section were taken on a single, well-controlled machine --- DKRZ Levante --- to keep the comparison against the original Fortran clean; results on other systems, including a different GPU vendor, are reported in Sect.~\ref{sec:xmach}, where the characteristics of each machine are also given.

On Levante, a GPU node carries four NVIDIA A100-80GB GPUs and two AMD EPYC-7763 processors; a CPU node carries two EPYC-7763 processors with 128 cores per node.

\subsection{Comparison across implementations}
The five implementations run an identical workload (Fig.~\ref{fig:compare}). The C port matches the original Fortran to within about 2\%, and is usually marginally the fastest: a literal C transliteration of Fortran ocean code is, to first order, performance-neutral. The Kokkos-CPU build carries a portability overhead of roughly 13--18\% over the C port, which is largest on the smallest mesh (where the per-rank work shrinks and the fixed per-rank infrastructure cannot exploit it) and nearly vanishes on the large meshes; the OpenMP back-end is a further $\sim$7--8\% slower than the pure-MPI Serial build.

  The CPU implementations strong-scale almost ideally --- 95--98\% efficiency per doubling on every mesh, with slight superlinearity on the small ones --- consistent with the FESOM2 scaling reported by \citet{koldunov2019scaling}, which stays close to linear down to roughly 500 surface vertices per core.  The GPU result is inverted over the course of the optimization campaign  (Section~\ref{sec:arc}). Node for node, the Kokkos-GPU build began the campaign roughly $3.8\times$ \emph{slower} than a CPU node and ended it between about $1.6$ and $3.7\times$ \emph{faster} on the high-resolution meshes, with the strongest speedup at low node counts, where each GPU is most fully loaded (Fig.~\ref{fig:ratio}). This advantage shrinks monotonically as nodes are added and the per-GPU subdomain thins. The crossover mesh is around the 0.64-million-vertices fArc grid: the 0.13-million-vertices CORE2 grid is GPU-favoured only at one or two nodes, while the multi-million vertices grids are GPU-favoured up to and beyond 16 nodes.

 \subsection{GPU profiling and the optimization}
  \label{sec:arc}
We did not aim for the best possible GPU performance. That would require rewriting the compute kernels, and probably changing the numerical code itself. Our aim was narrower: to see how far an LLM could speed up the code while keeping it bit-identical to the serial version. Every change described here meets that constraint.

When the whole model first ran on the GPU it was, node for node, about $3.8\times$ slower than a CPU node. Profiling showed why: the compute kernels used only about $7\%$ of each step, and most of the rest was spent copying whole fields between CPU and GPU. The cause was the halo exchange, which still moved each field back to the host, exchanged it there, and copied it to the device again. The step was limited by data movement, not by computation.

The first set of changes removed that movement. Over four steps the remaining halo fields were kept resident on the GPU, so that halos are exchanged directly between GPUs rather than through the host. At four GPU nodes this cut the step time from $16.3$ to $2.7\,$s, a sixfold gain, and reduced host--device traffic from about $13$ to $3\,$GB per step. The node-for-node comparison flipped from $3.8\times$ slower to $1.6\times$ faster.

The step was now limited by computation. A second set of bit-identical changes brought the same four-node case from $2.7$ to $1.27\,$s. The main kernels were restructured so that neighbouring GPU threads read neighbouring memory, and the column solver was changed to reuse its own scratch storage so that each column stays in fast memory. On the largest mesh these changes raised the node-for-node advantage from $1.6\times$ to about
$3.6\times$. The higher $3.7\times$ quoted earlier (Fig.~\ref{fig:ratio}) is not a further gain but the same build at its most favorable point --- the largest mesh on the fewest nodes --- and it shrinks as nodes are added and each GPU holds less work.

Two smaller changes helped only at the edges: removing transfers and halo exchanges that no kernel used, and, where the run becomes communication-bound at high node counts, merging paired halo messages into single ones (up to about $9\%$ there). Three changes did not help and were removed --- overlapping communication with computation, persistent MPI requests, and fusing kernel launches  --- because the time they targeted turned out to be small, or to come from load imbalance rather than communication. The recurring lesson was to measure before optimizing: the limiting factor moved at almost every step.

\begin{figure}[t]
  \centering
  \includegraphics[width=\linewidth]{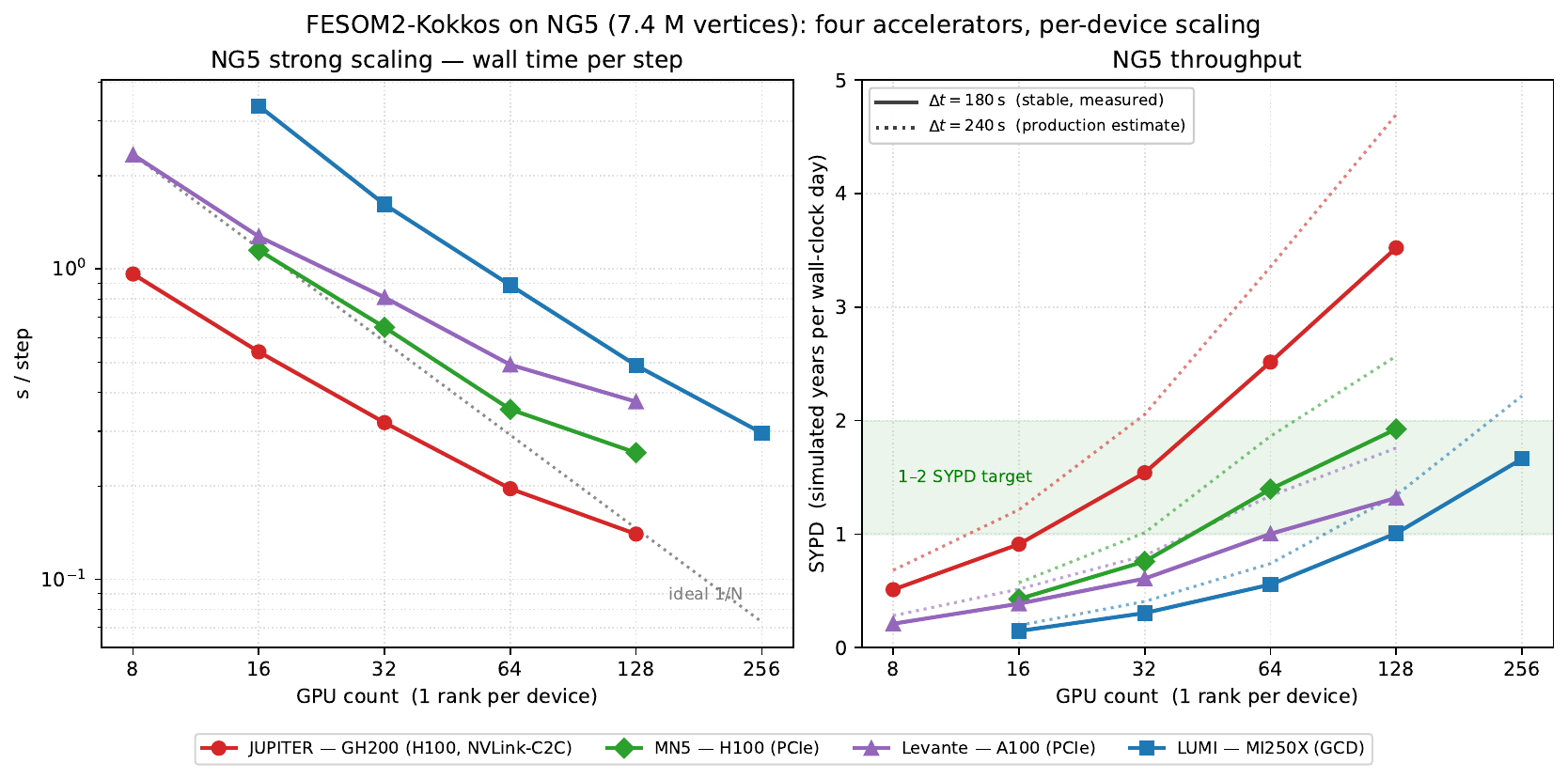}
  \caption{The single Kokkos source on the NG5 mesh (7.4 million surface vertices), run unchanged on four accelerators spanning both GPU vendors: NVIDIA GH200 (JUPITER), NVIDIA H100 (MareNostrum~5), NVIDIA A100 (Levante) and AMD MI250X (LUMI). The horizontal axis is the number of GPU devices, one
  MPI rank per device; for the MI250X one device is one compute die, so a LUMI node contributes eight. \emph{Left:} wall-clock time per step (strong scaling). \emph{Right:} throughput in simulated years per wall-clock day (SYPD); solid lines are measured at the stable time step ($\Delta t = 180\,$s), dotted lines are the $\Delta t = 240\,$s production-step estimate, and the shaded band marks the 1--2 SYPD production target. MareNostrum was timed at $\Delta t = 240\,$s, so its per-step values are slightly conservative relative to the others.}
  \label{fig:ng5_machines}
  \end{figure}
  
\subsection{Portability across machines}
  \label{sec:xmach}
The reason to write the port once, in Kokkos, is to run it on different hardware without touching the physics. We tested this on three EuroHPC systems, covering both GPU vendors. \emph{LUMI-G} (CSC) has nodes with four AMD MI250X cards and one 64-core AMD EPYC ``Trento'' CPU; each MI250X is two separate compute dies, so the runtime sees eight GPUs per node, and the AMD path uses the Kokkos HIP back-end. \emph{JUPITER Booster} (JSC) has nodes with four NVIDIA GH200 Grace--Hopper superchips, each joining an H100 GPU (96\,GB HBM3) to a 72-core Grace CPU over a 900\,GB/s NVLink-C2C link --- about $14\times$ the host--device bandwidth of a PCIe card --- and uses the CUDA back-end. \emph{MareNostrum~5 ACC} (BSC) has nodes with four PCIe-attached NVIDIA H100 (64\,GB) and two 40-core Intel Sapphire Rapids CPUs, also on the CUDA back-end. Levante, the machine used elsewhere in this paper, contributes its four PCIe NVIDIA A100 per node as the baseline.

Moving to each machine was straightforward. The two NVIDIA systems needed no source change at all: the existing CUDA build recompiled and ran. LUMI needed one small change. The device-residency layer of Section~\ref{sec:arc} had been written behind \texttt{\#ifdef KOKKOS\_ENABLE\_CUDA}, so on an AMD build it was silently skipped and the model fell back to moving every halo through the host --- correct, but slow. The fix was to replace that vendor-specific guard with a neutral one,

  \begin{verbatim}
  #if defined(KOKKOS_ENABLE_CUDA) || defined(KOKKOS_ENABLE_HIP)
  #  define FESOM_GPU_RESIDENT 1
  #else
  #  define FESOM_GPU_RESIDENT 0
  #endif
  \end{verbatim}

and to test \texttt{\#if FESOM\_GPU\_RESIDENT} at the roughly ten guarded sites instead of the CUDA macro. This turned the residency layer on for HIP without changing any arithmetic. It took less than a day. A single run confirmed it: with the residency layer switched on, the step time on NG5 dropped from $8.7$ to $3.4,$s

Figure~\ref{fig:ng5_machines} shows the outcome on the production-scale NG5 mesh (7.4 million surface vertices): strong scaling per GPU on the left, throughput on the right. The same source scales cleanly on all four machines. At a given number of GPUs the GH200 is the fastest device, about $2.7\times$ the A100; the PCIe H100 on MareNostrum is about $1.5\times$ the A100, a generational gain; and the MI250X is the slowest per compute die. Much of the GH200's lead is probably its NVLink-C2C link: the host--device wall that the residency campaign had to fight on Levante is largely absent there. For long production runs the relevant target is the 1--2 simulated-years-per-day (SYPD) band; at thirty-two nodes the GPU build reaches it on every machine, from about $1.3$ SYPD on the A100, $1.7$ on the MI250X and $1.9$ on the H100, to $3.5$ SYPD on the GH200.

\section{Discussion}
\label{sec:discussion}

\subsection{LLM-porting failure modes}
Most of the bugs we encountered during the FESOM2 port fall into a small number of classes that recurred across both stages and that we expect to appear in other scientific ports. They are worth setting out because each has a characteristic symptom and is caught by a specific level of the validation ladder.

The most expensive class was the latent bug: a term or constant that the configuration genuinely uses, mis-ported in a way whose effect is negligible at the short time step used for quick validation but becomes significant at the longer production step. The instance that cost the most was an Adams--Bashforth stabilization offset hardcoded as $10^{-9}$ where the Fortran value is $0.1$; the marginally-unstable scheme that resulted produced a central-Arctic instability only after about 110 model days, and only at the production time step, and its discovery was delayed for several sessions by a code comment that confidently stated the wrong value. Two checks now guard against this class. The first is to validate at the production configuration over one to two model years rather than at a short smoke test. The second is a standing rule that any statement about what the Fortran does must cite the exact file and line and the value actually read there, never a paraphrase; it was that rule which finally exposed the misleading comment.

A second high-impact class is specific to the move from shared to distributed memory: the halo loop-bound bug, in which a write loop that the Fortran runs over owned-plus-halo points is ported over the owned points only, leaving a stale halo. The result diverges silently, and only at some rank counts. It is caught cheaply by a probe that copies a field, exchanges the copy and reports any node that differs at the halo, and at the whole-model level by the requirement that the output be byte-identical across rank counts.

The remaining failure classes have lower impact, but they recur often enough to be worth naming, and each has its own detector. A field may be allocated and exchanged between processes yet never actually computed. It then stays at zero, and the comparison cannot detect the error, because the reference leaves the same field empty at the same phase. The mistake becomes visible only when a later phase first reads the field in a real computation. Until then, a cheap, always-on range check on fields with
known physical bounds flags the zero value directly. A tracer array may be indexed with the wrong vertical stride, taken from the loop bound instead of the array's allocated depth. This inflates a computed gradient by about a thousandfold and makes the solver fail within two steps; the per-kernel check catches it at once, far above the noise floor at the first step. Geometry stored per vertex but used per cell, or the reverse, corrupts layer thicknesses and control volumes, and the per-substep reference dump
locates it. A parameter set in the experiment's namelist may differ from the default written in the source, which the assistant tends to copy; the resulting systematic regional bias is caught by the multi-year comparison against the reference. In every case the lesson is the same: these failures would break an unguarded port, but each is caught cheaply by a specific level of the validation rather than by manual inspection.

\subsection{What the LLM did well and badly}
The LLM was good at exactly the parts that a human finds hardest to sustain over a long project: tireless line-by-line translation, building and maintaining heavy validation harnesses, applying mechanical code transformations that are type-exact by construction, carrying lessons across many sessions through the memory store, and profiling-driven optimization. Its characteristic weaknesses were: a standing tendency to ``simplify'' or ``improve'' unless repeatedly constrained, and a willingness to state what the Fortran does from a plausible paraphrase rather than from the line itself. Both are directly addressed by the practices in Section~\ref{sec:methods}. Humans remained essential throughout as domain experts who set strategy, point to possible direction to resolve issues, reviewed the work, and recognized the subtle physical and validation errors that the gates flagged but did not diagnose.

\subsection{Future work}

This study should be read as a strong worked example rather than as a controlled experiment. We ported one production ocean--sea-ice model, with one team, one principal agentic LLM assistant, and one main target programming model. The result demonstrates that an LLM-assisted port can preserve the physics of a full geophysical model when constrained by a literal-translation rule and a tiered validation ladder, but it does not by itself prove that the same workflow is optimal for all models, all LLMs, or all target architectures.

A first direction for future work is methodological. The most important comparison is between direct and staged translation paths. In this work we used a Fortran$\rightarrow$C$\rightarrow$C++/Kokkos route because the C stage separated numerical translation from parallelization, collapsed the configurable Fortran source onto the active model configuration, and gave the Kokkos port a deterministic reference. However, the intermediate stage also has a cost: it creates another code base to build, validate, document and maintain. A direct Fortran$\rightarrow$C++/Kokkos translation would be more attractive if it could be made equally reliable, because it would avoid that intermediate representation and shorten the path to the final performance-portable implementation. A controlled study should therefore compare at least three workflows: direct Fortran$\rightarrow$C++/Kokkos, Fortran$\rightarrow$C++ followed by Kokkos parallelization, and the staged Fortran$\rightarrow$C$\rightarrow$C++/Kokkos route used here. Such a comparison should be run across several LLMs, with the same validation ladder and the same production configuration, so that differences in success rate, debugging effort, fidelity and final performance can be attributed to the workflow rather than to experience with a single model.

The second methodological step is to apply the recipe to other geophysical models. The reusable part of the present work is not the FESOM2-specific code, but the procedure: restrict the first target to one exercised configuration, translate literally, keep a deterministic reference, port kernels incrementally, validate each kernel against its twin, and require multi-year agreement at the production configuration. Applying this recipe to other ocean, atmosphere, sea-ice or coupled-model components would test its general validity. In particular, models with structured grids, different halo patterns, different vertical coordinates, or different solver structures may expose different LLM failure modes. Repeating the exercise across such models would turn the present case study into a more general account of when LLM-assisted scientific-code ports are trustworthy, when they fail, and which parts of the validation ladder are essential.

A second class of future work concerns performance and energy efficiency. The largest unused performance lever is mixed precision. The present port keeps the numerics in double precision in order to preserve the bit-identical Serial reference and make debugging exact. That choice is conservative and appropriate for establishing fidelity, but it is unlikely to be optimal for production GPU runs. Many ocean-model kernels are limited by memory traffic rather than by floating-point throughput. In such kernels, using single precision or carefully selected mixed precision can reduce the number of bytes moved per grid point, reduce pressure on memory bandwidth, and increase arithmetic intensity. This is especially relevant on modern GPU systems, including tightly coupled CPU--GPU nodes such as NVIDIA Grace Hopper systems, where high bandwidth and large memory capacity make data placement and data movement central performance questions. A future mixed-precision version should therefore be introduced selectively rather than globally: prognostic variables, reductions, solvers and equation-of-state calculations may need different precision choices, and each choice must be justified by multi-year climate statistics rather than by short numerical tests alone.

Performance should also be evaluated by energy-to-solution, not only by time-to-solution. Climate simulations are long-running workloads, and their scientific value has to be weighed against their computational cost and carbon footprint. A GPU port that reaches the required simulated-years-per-day target is useful, but a production climate model should also minimize the energy required per simulated year. Future benchmarking should therefore report Joules per simulated year, or an equivalent energy-to-solution metric, alongside wall-clock throughput and scaling. Such measurements may also change optimization priorities: the fastest configuration is not necessarily the most energy-efficient one, especially when communication, host--device transfers, or low GPU occupancy dominate at high node counts.

A third set of limitations is specific to the present FESOM2 port. The ported configuration is sufficient to demonstrate the method and to run the fidelity and performance experiments reported here, but it is not yet a complete production replacement for the full Fortran model. For production use, the port needs restart input/output, support for the remaining vertical-coordinate variants, and the additional physics packages used in coupled climate configurations. Extending the port to these components would therefore provide a stronger test of the general porting recipe.

\section{Conclusions}
\label{sec:conclusions}
To our knowledge, this is the first time a real, full-physics ocean model has been ported from Fortran to GPUs with a large language model and verified, by direct measurement, to preserve its physics. Working under expert direction, an agentic LLM translated FESOM2 --- about 74{,}000 lines of Fortran for the ocean and sea ice --- first into C and then into a single C++/Kokkos source that runs unchanged on CPUs and on GPUs from both major vendors: NVIDIA through CUDA and AMD through HIP. At each stage,
acceptance required passing validation. The C port reproduces the original Fortran's ocean over five model years, with a five-year-mean sea-surface-temperature difference of $0.006\,^{\circ}$C and no drift in the deep ocean. The Kokkos Serial and OpenMP back-ends reproduce that C reference byte-for-byte over a full simulated year. On the GPU the model stays statistically close: its residual difference is about three orders of magnitude below the C-versus-Fortran uncertainty inherited from the first stage. The same source reaches the 1--2 simulated-years-per-day production band on multi-million-vertex meshes on every machine tested --- NVIDIA A100, H100 and GH200, and AMD MI250X --- and up to 3.5 simulated years per day on the GH200.

What made the port reliable was not any single tool but a discipline: translate literally, and hold each stage to the acceptance bar that fits it --- multi-year statistical agreement where byte-level equality is neither attainable nor meaningful, and byte-level identity where it is. This tiered validation is what makes an LLM-assisted port trustworthy, because it converts the model's most dangerous failures --- silent physics errors that compile and run cleanly --- into immediate, localized failures, each caught by a specific level of the validation. 

The reusable result is not only the FESOM2 code, but also this procedure, and it is small enough to reproduce, and to extend --- to mixed precision and energy-to-solution, and to other ocean, atmosphere and coupled models. We offer it as a practical path for bringing the large base of Fortran geophysical models into modern, well-supported languages, and onto the accelerators that climate prediction increasingly depends on.

\section*{Code and data availability}

The Fortran reference is FESOM2 v2.7.3 with the C-port reference-dump instrumentation, tag \texttt{fesom2.7.3-cport-instr} (commit \texttt{9271ae92}), available at \url{https://github.com/FESOM/fesom2}; it was built with Intel oneAPI 2022.0.1 and OpenMPI 4.1.2 on Levante. The C port is available at \url{https://github.com/koldunovn/fesom_port}, and the C++/Kokkos port at \url{https://github.com/koldunovn/fesom_kokkos}. The input data required to run the models --- meshes, initial conditions, and atmospheric forcing --- are available from the authors on request.
  

\section*{Competing interests}
The authors declare no competing interests.

\section*{Acknowledgements}
This work was primarily supported through the core funding of the Alfred Wegener Institute, Helmholtz Centre for Polar and Marine Research (AWI), within the Helmholtz Association. Additional support for individual authors was provided through the following projects and funding sources. The contribution by NK was supported by the TerraDT - Digital Twin of Earth system for Cryosphere, Land surface and related interactions project, which has received funding from the European Union’s Horizon Europe research and innovation programme under Grant Agreement no.101187992. SC is funded by Warm World:Better project funded by the German Federal Ministry of Research, Technology and Space under the funding code 01LK2202A. The responsibility for the content of this publication lies with the author. SD and NK were supported by projects M5 and S1 of the Collaborative Research Centre TRR181 “Energy Transfer in Atmosphere and Ocean” funded by the Deutsche Forschungsgemeinschaft (DFG, German Research Foundation)—Projektnummer 274762653. The work of SB has been supported by the European Union’s Destination Earth Initiative and relates to tasks entrusted by the European Union to the European Centre for Medium-Range Weather Forecasts implementing part of this Initiative with funding by European Union. Views and opinions expressed are those of the authors only and do not necessarily reflect those of the European Union or the European Commission. Neither the European Union nor the European Commission can be held responsible for them.  TJ was supported by the EU funded EERIE project, which received funding from the European Union’s Horizon Europe research and innovation programme under Grant 1173 Agreement No. 101081383. 

The authors thank the German Climate Computing Center (DKRZ), Jülich Supercomputing Centre (JSC), Barcelona Supercomputing Center (BSC) and IT Center for Science (CSC) for access to their computing infrastructure. We also thank the authors of the GitHib `umputun/cc-thingz` repository for their excellent set of Claude Code skills.

\bibliography{refs}
\newpage
\appendix
\section{The Fortran-to-C port: order of work and main difficulties}
\label{sec:appendix-cport}

This appendix records, as a companion to Sect.~\ref{sec:methods}, the order in which the C
reference was built and the problems that took the most time to solve. It is reconstructed from
the port's own plan files and project-memory store. The content is specific to FESOM2; the
general practices are in Section~\ref{sec:methods} and the cross-cutting failure-mode catalogue in
Section~\ref{sec:discussion}.

\subsection{Order of work}
The port grew in phases. Each phase ended in a documented, validated state, and wherever a phase
added new physics it also added a switch that returns that physics to a no-op, so the phase
boundary could be checked byte-for-byte against the previous state.

The first phases brought up the ocean on a single rank.
\begin{enumerate}
\item \emph{Serial baseline.} Mesh input, areas, edge lengths and scalar gradients; state
  allocation with structures mirroring the Fortran modules; a constant-temperature,
  constant-salinity initial state with no forcing; the linear free surface only; the full
  nonlinear equation of state and hydrostatic pressure; the conjugate-gradient
  sea-surface-height solver; second-order Adams--Bashforth, the pressure-gradient force,
  explicit vertical viscosity and bottom drag; upwind tracer advection; and
  Pacanowski--Philander vertical mixing. Acceptance: a rest state stays at rest, and a
  sea-surface-height bump propagates as a gravity wave at the expected speed.
\item \emph{Idealized mesh.} Flux-corrected tracer advection (with care for the sign of the
  vertical advective flux), horizontal viscosity with backscatter, the implicit/explicit
  vertical-velocity split, and analytical forcing. Acceptance: a thousand short steps without
  instability.
\item \emph{Realistic single rank (CORE2).} The PHC initial state (horizontal interpolation, the
  neighbour-fill of dry points, vertical filling, and the in-situ-to-potential temperature
  conversion), the CORE2 mesh specifics (rotation detection, clockwise element orientation,
  partial cells), the level vertical coordinate, the JRA55 atmospheric reader with bulk formulae,
  and sea-surface-salinity restoring with runoff.
\end{enumerate}
The first two of these phases ran almost entirely on FESOM2's small idealized test mesh, the
  \texttt{pi} mesh: a global triangular mesh of \num{3{,}140} surface nodes, \num{5{,}839} elements
  and about \num{23} vertical levels that completes a thousand time steps in under a minute on one or
  two cores. That speed made it the workhorse of the early port, where the dynamical core was rebuilt
  one kernel at a time and had to be re-exercised many times a day. The production model drives this
  mesh from atmospheric reanalysis read through a netCDF interface, but that reader did not yet exist
  at this stage; rather than hold up the bring-up on input and output, we had the assistant synthesise
  the forcing in code. The result is a self-contained analytical wind field --- a surface wind stress
  that varies as a cosine of latitude, giving easterly trades at low latitudes and westerlies at
  mid-latitudes, with no heat or freshwater flux beyond an optional relaxation of sea-surface
  temperature. This is enough to drive surface Ekman transport and a gyre-scale circulation beneath
  it, which is exactly the behaviour needed to test tracer advection, horizontal viscosity and the
  vertical-velocity split for stability over the thousand-step acceptance run. One subtlety the
  assistant handled correctly is that the \texttt{pi} mesh is geographically rotated, so the wind had
  to be built from the true, un-rotated latitude of each element rather than from the model's own
  coordinates, or the pattern would not have aligned with the equator. The idealized forcing was
  retired once the realistic CORE2 phase introduced the netCDF reader and bulk formulae; its purpose
  was to separate validation of the numerics from the input/output system during the period when both
  were being built at once.
The middle phases added parallelism and the missing physics.
\begin{enumerate}\setcounter{enumi}{3}
\item \emph{Distributed memory (MPI).} The partition reader, a generic halo exchange with
  explicit pack and unpack buffers, a halo-identity self-test, allocation of every per-node and
  per-element array at owned-plus-halo size, an audit of every loop bound against the Fortran,
  halo exchange after every kernel whose output another rank reads, the parallel
  conjugate-gradient solver, and a gather-to-rank-0 snapshot writer. This is where most of the
  hard bugs lived (below).
\item \emph{Sea ice.} Built in order: a scaffold (state and a no-op driver), the thermodynamics,
  the ice-to-ocean coupling (heat, water and salt fluxes), the elastic--viscous--plastic
  dynamics with its 120-subcycle solver, the ice-aware surface stress, and finally the
  flux-corrected advection of the ice variables. Sea ice stoped a long-run
  cold drift of the high-latitude surface ocean; acceptance was that the minimum temperature
  stays near the freezing point over a simulated month.
\item \emph{Mesoscale eddy parameterization (GM/Redi).} Built as a sequence of small, separately
  validated steps from the mesh prerequisites through the slope and neutral-slope computations,
  the streamfunction solve, the bolus velocities and the rotated diffusion, ending with a master
  no-op switch whose ``on'' state byte-matches the pre-GM model.
\end{enumerate}

Alongside these we built an output subsystem. This is the one place where we deliberately did not
port the Fortran: rather than transliterate FESOM2's output code we designed a smaller,
configuration-driven writer that emits per-variable daily, monthly or yearly time means to
readable netCDF, with room for a future zarr backend.

A second wave of work, in later sessions, raised the model to the production configuration and
closed the remaining differences from the Fortran.
\begin{enumerate}\setcounter{enumi}{6}
\item \emph{Production time step.} Raising the step from 500 to 1800 seconds exposed a halo bug
  that had been harmless at the shorter step (the stale-coefficient bug, below).
\item \emph{Closing residual biases.} Two physics terms that the early port had simplified were
  restored as faithful ports once differential testing localized the bias each caused: the
  third-order (MFCT) horizontal tracer advection, replacing an interim central scheme that was
  too diffusive at sharp fronts and left a river-mouth salinity bias and an equatorial
  cold-tongue bias; and shortwave penetration through the upper column, replacing a single-layer
  heat input that had left the surface too warm and the layers below too cold.
\item \emph{KPP vertical mixing.} The original FESOM K-profile scheme was ported in a sequence of
  small steps --- scaffold, the boundary-layer-depth routine as the highest-risk piece, the
  boundary-layer coefficients, the enhancement and assembly, wiring it as a selectable scheme,
  and multi-year validation --- and then made the default. The branches the coordinated-experiment
  configuration does not enable (double diffusion, the non-local flux) were ported as gated-off
  stubs rather than full bodies.
\item \emph{Multi-year validation.} Two- and five-year runs against the matched Fortran code confirmed
  the bounded drift and long-term agreement reported in Section~\ref{sec:fidelity}.
\end{enumerate}

\subsection{The problems that took the most time}
A small number of problem classes accounted for most of the lost time.

\emph{Halo write-loop coverage} was the highest-impact class. Many Fortran write loops run over
owned-plus-halo points precisely so that a later exchange can be skipped; the bound looks unusual
but matters, and a port that writes only the owned points leaves the halo stale. The bug is
silent on one rank and appears at some rank counts and not others. The hardest single instance
was the one that capped the time step: the bulk-formula routine computed two atmosphere--ocean
exchange coefficients over owned points only, while sea-ice thermodynamics reads them at the
halo. The stale halo coefficients made the sea-ice state diverge slightly across ranks, which
made the redundantly-computed wind stress on shared boundary elements differ between ranks, which
broke the cancellation in the sea-surface-height right-hand side, which made the free surface
non-conservative. Because the resulting instability grows with the square of the time step, it
was tolerable at 500 seconds and fatal at 1800. The Fortran code was stable at 1800, which under the
literal-translation rule proved the fault was in the port, not the physics. The tool that found
it copies each candidate input field, exchanges the copy, and reports any field that differs from
the original at the halo; run over the whole candidate list at once, it named the two coefficients
directly.

\emph{Array size versus a reader's loop bound} is the cross-module form of the same problem: one
module allocates a per-node array at owned size only, and another reads it in an owned-plus-halo
loop, reading uninitialized memory at the halo. A sea-ice run crashed at the second step only on
multiple ranks; an always-on cheap probe printed an ice thickness of seven million metres at a
halo node where the air temperature read as zero Kelvin, naming the cause at once. The resulting
rule is to size every per-node field that any downstream routine might read at the halo at
owned-plus-halo, since the memory cost is trivial and the failure is severe.

\emph{Tracer storage stride.} Tracers are mid-layer values, but their storage uses the full
layer-interface count as the stride; indexing with the smaller count shifts every read by one
slot per vertex. The symptom was a derived slope a thousand times too large, a bolus velocity of
twenty metres per second, a salinity overshoot to about sixty PSU at the
first step, and a solver failure at the second. The rule is to take the stride from the
allocation shape, never from the iteration range.

\emph{Producers ported after their consumers.} An auxiliary field is sometimes allocated and
exchanged before the Fortran routine that fills it has been ported; it then sits at zero until a
later phase introduces its first reader, and the phase looks bit-identical to a baseline that also
left the field empty. The rule is that when a phase introduces a reader of an auxiliary field, the
port is searched for a writer of that field, and if none exists the producer is ported first.

A separate group of difficulties came not from the port but from FESOM2 itself --- parts that
mislead any reader, in any language. The model stores one depth per node but builds its geometry
per cell, so confusing the two corrupts layer thicknesses and control volumes. The PHC initial
file holds in-situ temperature while the model uses potential temperature, and skipping the
conversion leaves a silent deep-water bias. Element node order must be normalized to clockwise at
start-up, or half the solver terms take the wrong sign. Runoff enters the freshwater flux through
two contradictory conventions depending on which thermodynamics path is active, inviting a
double count. The KPP scheme exists in two different implementations selected by one
namelist option, and the coordinated experiment uses the wrong one. And two effects look like
port bugs but are accepted Fortran behaviour and must not be chased: the neighbour-fill of dry
points and the MPI sum reductions both depend on the partition, so a few coastal nodes and the
solver diagnostics differ by small amounts between rank counts.

Finally, the earlier port had failed on a set of well-meant simplifications that we now knew
to avoid: linearizing the thermal and haline expansion coefficients (which broke the
boundary-layer depth), adding a depth cap the Fortran does not have, replacing a lookup table with
a closed form, and mismatching the Adams--Bashforth history length to its order (which inflated
the Coriolis weight and made the surface height run away). Each had produced plausible output and
a multi-week search; together they are why the literal-translation rule exists.

\section{The C-to-Kokkos port: order of work and main difficulties}
\label{sec:appendix-kokkos}

This appendix is the detailed record behind Sections~\ref{sec:methods}--\ref{sec:performance}: the
order in which the Kokkos port was built and the problems that took the most time. The summarized
results are in the main text; here we give the milestone-by-milestone account and some mechanism
detail there was no room for. As in Appendix~\ref{sec:appendix-cport}, it is reconstructed from
the port's plan files and project-memory store.

\subsection{Order of work}
The port was built in five phases. The order is strict: each device-kernel phase rests on the
data layer and the per-kernel Serial check, and each performance phase rests on the whole model
already being device-resident and validated over multiple years.

The quick checks that gated each of these kernels ran on the same small \texttt{pi} mesh that had brought up the C port (Appendix~\ref{sec:appendix-cport}) --- the idealized test mesh referred to below --- which verifies a kernel in seconds, while the realistic, ice-active configuration was kept for longer fidelity gates.
  
\emph{Phase 1 --- language flip.} All of the model's roughly forty translation units were
compiled as C++ with no change of algorithm. The only mechanical obstacle was that C++ does not
allow the implicit \texttt{void*}-to-typed-pointer conversion that C permits at every allocation
site; a script inserted type-exact casts at all 305 sites. With no device kernels yet present, the Serial, OpenMP and CUDA builds were all bit-identical to the C reference --- the flip changes no arithmetic, which is what makes the rest of the port safe to do one kernel at a time.

\emph{Phase 2 --- data layer.} Each of the 126 persistent arrays was wrapped in a Kokkos
\texttt{DualView}. The migration was made bit-identity-preserving by a single discipline: the
\texttt{DualView} owns the storage, and the original raw pointer is kept as a non-owning alias to
its host copy, set once after allocation, so that the several hundred existing indexed accesses
and the indexing macros stay byte-for-byte unchanged. The synchronization policy is
host-authoritative and lazy, so the data layer by itself moves no data
between host and device; a field migrates only when the kernel that consumes it is ported. The
contract --- which memory space is authoritative for each field at each point in the step --- is
written out as a sync map and made executable by an optional build that aborts if the host ever
reads a field whose device copy is the authoritative one. Acceptance for the phase was a one-year
Serial \emph{and} OpenMP run bit-identical to the C reference, with that guard provably unable to
fire, which is itself the proof that the whole model is still uniformly host-authoritative.

\emph{Phase 3 --- kernels to the device.} Kernels were moved one at a time, each behind the
per-kernel check (Section~\ref{sec:ladder}), in dependency order: the equation of state first (the
first device arithmetic, and the point at which CUDA stops being bit-identical), then
Pacanowski--Philander mixing, the K-profile scheme (the largest single kernel, about a thousand
lines), the pressure-gradient and momentum terms (which introduce the first scatter), the
vertical-coordinate and eddy-parameterization kernels, the flux-corrected tracer advection (three
scatters), and the implicit vertical tracer diffusion. The sea-surface-height solver and then the
entire sea-ice step (the elastic--viscous--plastic rheology, ice advection, thermodynamics and the
ice--ocean fluxes) followed, together with the rank-to-device mapping for multiple GPUs. Acceptance
for this phase was the whole coupled model running device-resident with a one-year Serial result
still byte-identical to the C reference --- that is, the device-targeted kernels compose
bit-identically over a full year, not merely one at a time.

\emph{Phase 4 --- performance.} This was the bulk of the project, a long loop of profile, change,
re-profile. It moved the halo exchanges onto the device with GPU-aware MPI, then --- guided by
profiling at the production mesh --- made the large fields device-resident so that the step stopped
being limited by data movement over the PCIe bus, and finally worked the communication and compute
limits that remained. 

\emph{Phase 5 --- multi-year validation and scaling, interleaved.} Every performance milestone
that touched the device path was closed with a one-year realistic run compared against both the C
port (which isolates the GPU reduction and scatter drift) and the Fortran (the absolute budget),
so that no optimization could quietly change the science. The final strong-scaling and
throughput sweep, on meshes up to 7.4~million surface nodes and up to 32 GPU nodes, produced the
numbers in Section ~\ref{sec:performance}.

\subsection{The problems that took the most time}
Three of the difficulties below are the sources of CPU--GPU divergence; here we give the detail and say at which milestone each arose. The others are specific to making the device port fast and correct at scale.

\emph{Coherence of the data layer.} The risk in wrapping 126 arrays in a host/device container is
not the wrapper but the surrounding C-style code. Clearing a structure that now holds such a
container with \texttt{memset}, or copying it with \texttt{memcpy}, is no longer defined
behaviour; and a host write through the kept raw alias is invisible to the container's
change-tracking, so a later device synchronization silently copies stale data. The
storage-owned-by-the-container discipline above removes the first class of error, and the
host-authoritative policy plus the executable sync-check build removes the second by making any
violation abort rather than corrupt a field quietly.

\emph{The equation-of-state first-divergence.} Moving the equation of state to the device ends
bit-identity for CUDA, with the small, stable signature reported in Sect.~\ref{sec:fidelity}. The
work here was to characterize that signature precisely --- in-situ density near
$3\times10^{-12}$, buoyancy frequency near $3\times10^{-15}$, both steady across steps, with only
the depth-integrated velocity drifting --- and record it, so that a real regression (a growing
density difference, or mixing differences that spread rather than staying at a few isolated
marginally-stable nodes) is told apart from expected rounding at a glance. The host build keeps
fused multiply-add off, so the host and Serial-device results stay bit-identical even though
this is a code-generation no-op on the test machine; it is kept as the explicit, portable
determinism setting.

\emph{The scatter trade-off.} The kernels that accumulate edge or element quantities into shared
nodes are floating-point sums whose result depends on the order of addition, and the C reference
fixes that order through its global edge loop. We port them with atomic addition in that same
global order. On the Serial back-end the loop runs on one thread, so each atomic add is an
ordinary, unfused add and the result is byte-for-byte the C reference, which is what lets the
per-kernel check hold at exactly zero. On the threaded and GPU back-ends the additions race and
reassociate, giving the small per-step floor of Section ~\ref{sec:fidelity}. This is why the OpenMP
back-end, bit-identical for the pure map kernels, is \emph{not} bit-identical for these. It is
also why the trade-off is fundamental rather than a tuning choice: rewriting a scatter as a gather
(summing each receiving node's edges in its own order) would restore threaded bit-identity only by
changing the association away from the global edge order, which would break the Serial check that
the whole scheme rests on.

\emph{The stale host-copy bug and how it was found.} After the halo exchanges were moved onto the device, one field was left device-authoritative while a remaining host routine read its stale host copy. It was invisible on
the idealized test mesh for several commits and surfaced only on the realistic, ice-active
configuration, where it grew to about $0.4$ against a $10^{-3}$ floor. Finding it was hard because
the obvious test --- toggle each candidate field's synchronization and see which matters --- is
confounded on a chaotic system: removing a synchronization also removes a device fence, which by
itself reorders the scatters enough to move the trajectory. The discriminator that worked was to
keep every synchronization in place (so device scheduling is unchanged) and instead overwrite each
host copy with not-a-number \emph{after} it, without marking the host modified, so that the device
stays correct and only a genuine host read sees the poison. Three of the four suspected fields
turned out to be read only on the device; poisoning them changed nothing. The fourth, the node
velocity consumed by the surface wind-stress formula, produced an immediate failure --- it was the
one real host reader, and the reason the bug could only appear with active sea ice. The episode is
why the twenty-step gate on the realistic, ice-active configuration is mandatory before any commit
that touches device data movement.

\emph{The device-residency campaign.} After the whole model was on the device, the production mesh
was, surprisingly, slower than the CPU node for node. A CUDA timeline trace settled the cause: the
device kernels occupied only about 7\% of the step and roughly three-quarters of it was spent
moving whole fields across the PCIe bus, because each three-dimensional halo exchange still staged
its field to the host. The fix was applied over four milestones --- making the eddy-parameterization
and vertical-coordinate halo fields device-resident, then the salinity and density fields (which
required moving a host-side salinity floor onto the device), then finishing the
eddy-parameterization chain, and finally moving the surface bulk-formula computation onto the
device (which both removes a single-threaded host kernel and the transfer it forced). The step
went from being limited by data movement to being limited by computation, the host--device traffic
fell several-fold, and the node-for-node comparison crossed from the GPU being far slower to the
GPU being faster. The lesson recorded alongside it is that a per-phase wall-clock timer cannot separate computation from data movement within a phase, and that profiling must be done at the production mesh, not the fast development mesh.

\emph{Reproducing the communication-bound regime cheaply.} Once data movement was no longer the
limit, the limit at scale moved to communication, which is governed by the number of surface
points per rank rather than by absolute mesh size. The expensive sixteen-node target on the
largest mesh could therefore be reproduced on eight nodes of an intermediate mesh at the same
points per rank, letting the communication experiments run in minutes. The proxy is faithful for the halo and geometry terms but not for the solver's iteration count, which is set by the conditioning of the particular mesh, so the solver fraction was always measured on the mesh itself.

\begin{table*}[t]\centering
  \caption{The GPU optimization ledger on the production mesh (NG5, 7.4\,million surface nodes, 70
  levels) at four GPU nodes (\texttt{dist\_16}), Levante. Each row is one milestone; \emph{s/step} is
  the clean same-node whole-step time and $\Delta$ the per-milestone change from a controlled
  same-allocation test. The campaign moves the node-for-node comparison against a CPU node from
  $3.8\times$ slower to about $3.6\times$ faster. A dash in the \emph{s/step} column marks milestones
  for which only the per-step $\Delta$ was recorded, not a separate whole-step timing. All
  optimizations are bit-identical on the deterministic (Serial) back-end.}
  \label{tab:ledger}
  \begin{tabular}{llp{8.2cm}rr}
  \toprule
  Milestone & Lever & Change & s/step & $\Delta$ \\
  \midrule
  M5.12 & ---         & Whole model on device (campaign start; $3.8\times$ slower than a CPU node) & 16.27 & --- \\
  \midrule
  M5.13 & residency   & Three-dimensional halo fields made device-resident                       & 6.12  & $-62\%$ \\
  M5.14 & residency   & Salinity, density and vertical-velocity fields made device-resident (parity crossed) & 3.80 & $-38\%$ \\
  M5.15 & residency   & Eddy-parameterization fields made device-resident; verify-only syncs removed & 3.46 & $-9\%$ \\
  M5.16 & residency   & Surface bulk-flux formula moved to a device kernel ($1.6\times$ faster)   & 2.68  & $-22\%$ \\
  \midrule
  M5.17 & (reverted)  & Overlap communication with computation --- wait is load imbalance, not recoverable & --- & --- \\
  M5.18 & coalescing  & Tracer smoother re-parallelized for coalesced memory access                & ---   & $-14\%$ \\
  M5.19 & coalescing  & Velocity-tendency and eddy-parameterization kernels coalesced              & ---   & $-5\%$ \\
  M5.20 & residency   & Two further large fields made device-resident                              & ---   & $-18\%$ \\
  M5.21 & coalescing  & Transport (flux-corrected) kernels coalesced; a redundant transfer removed & 1.77  & $-8\%\,/\,-3\%$ \\
  M5.22 & coalescing  & Remaining transport sub-kernels coalesced ($2.95\times$ faster)            & 1.47  & $-3\%$ \\
  \midrule
  M5.23 & comm.\      & Sea-ice halo exchanges fused into single messages\textsuperscript{a}       & 1.47  & $-9\%$\textsuperscript{a} \\
  M5.24 & cache       & Tridiagonal solves done in place, smaller working set ($3.6\times$ faster) & 1.27  & $-2\%$ \\
  \bottomrule
  \end{tabular}

  \vspace{2pt}
  {\footnotesize \textsuperscript{a}\,Flat at four nodes (compute-bound); the $-9\%$ is reached only in
  the communication-bound regime at high node count, where the message-count reduction pays.}
  \end{table*}
\end{document}